\documentstyle{article}
\begin{document}
\date{}
\title{Jordan-Schwinger map, 3D harmonic oscillator constants of motion, and
classical and quantum parameters characterizing electromagnetic wave
polarization}
\author{R D Mota$^{1}\footnote{This work was prepared during my posdoctoral
year at Departamento de Matem\'aticas del Centro de Investigaci\'on
y Estudios Avanzados del IPN, M\'exico D F, 07000, Mexico.}$, M A
Xicot\'encatl$^{1}$ and V D Granados$^{2}$}

\maketitle

\begin{minipage}{0.9\textwidth}
\small $^{1}$ Unidad Profesional Interdisciplinaria de
Ingenier\'{\i}a y Tecnolog\'{\i}as Avanzadas, IPN. Av. Instituto
Polit\'ecnico Nacional 2580, Col. La Laguna Ticom\'an, Delegaci\'on
Gustavo A. Madero,
07340 M\'exico D. F., Mexico.\\
\linebreak $^{2}$ Escuela Superior de F\'{\i}sica y Matem\'aticas,
Instituto Polit\'ecnico Nacional,
Ed. 9, Unidad Profesional Adolfo L\'opez Mateos, 07738 M\'exico D F, Mexico.\\
\end{minipage}
\rm

E-mail: mota@esfm.ipn.mx, xico@math.cinvestav.mx\\

\begin{abstract}

In this work we introduce a generalization of the Jauch and
Rohrlich quantum Stokes operators when the arrival direction from
the source is unknown {\it a priori}. We define the generalized
Stokes operators as the Jordan-Schwinger map of a triplet of
harmonic oscillators with the Gell-Mann and Ne'eman $SU(3)$ symmetry
group matrices. We show that the elements of the Jordan-Schwinger
map are the constants of motion of the three-dimensional isotropic
harmonic oscillator. Also, we show that generalized Stokes Operators
together with the Gell-Mann and Ne'eman matrices may be used to
expand the polarization density matrix. By taking the expectation
value of the Stokes operators in a three-mode coherent state of the
electromagnetic field, we obtain the corresponding generalized
classical Stokes parameters. Finally, by means of the constants of
motion of the classical three-dimensional isotropic harmonic
oscillator we describe the geometric properties of the polarization
ellipse
\end{abstract}

PACS numbers: 42.50.-p, 42.25.-p, 42.25.Ja, 11.30.-j, 03.65.Fd

\section{Introduction}

In both classical and quantum optics, Stokes parameters have proven
to be intuitive and practical tools for characterizing the
polarization state of light [1-6].

A classical or quantum electromagnetic wave propagates, in general,
in an arbitrary but fixed direction in space. However for the study
of the polarization properties of the wave, the knowledge of the
propagation direction of the wave allows us to chose a coordinate
system in such away that the propagation is along the $z$ axis and
the polarization vector rests on the $x-y$ plane (\it i. e. \rm it
has only two polarization components) \cite{Jackson}. Also, the
knowledge of the propagation direction allows to use a 2D apparatus
(polarizers, wave plate rotators, etc.) placed perpendicular to the
wave propagation direction to measure the polarization
characteristics of the wave. The works [1-9] were done under the
assumption that the arrival direction from the source of the
electromagnetic wave was known. In \cite{Stokes} G. Stokes studied
the polarization properties of a quasi-monochromatic plane wave of
light in an arbitrary polarization state by introducing four
quantities, known since then as the Stokes parameters. Wiener used
the $2\times2$ unit matrix and the Pauli matrices as a basis to
expand the coherence tensor \cite{Wiener}. Fano \cite{Fano} showed
that the Stokes parameters are the expansion coefficients of the
coherence tensor. Stokes parameters obtained under an {\it a priori}
knowledge of the propagation direction will be refered in this work
as the usual classical or quantum Stokes parameters, and they are
well described in references \cite{Jackson,Simons} and \cite{Jauch},
respectively.

When we do not known {\it a priori} the propagation direction of the
wave, we already do not get the above adequate chose of the
coordinate system, so that, in general, the three components of the
polarization vector are nonzero. In this case, the three-dimensional
coherence tensor must be used to obtain a complete polarization
characterization \cite{Roman,Suizos}. Roman \cite{Roman} used the
basis of nine $3\times 3$ matrices which form the Kemmer algebra to
define the generalized Stokes parameters as the expansion
coefficients of the correlation matrix. In \cite{Suizos} Carozzi
{\it et. al.} defined the generalized Stokes parameters as the
expansion coefficients of the spectral density tensor in terms of
the $SU(3)$ Gell-Mann and Ne'eman matrices.

In this work we introduce a generalization of the Jauch and Rohrlich
quantum Stokes operators when the arrival direction from the source
is unknown {\it a priori}. For simplicity we study the case of a
monochromatic quantized plane electromagnetic wave that propagates
in a fixed but arbitrary direction in space. Also, we will use
$\hbar=\omega=\mu=1$, where $\mu$ is the mass of each 1D harmonic
oscillator and $\omega$ is the angular frequency of either the
electromagnetic wave or each harmonic oscillator. In section 2, we
define the generalized quantum Stokes operators as the
Jordan-Schwinger map of a triplet of  harmonic oscillators with the
Gell-Mann and Ne'eman $\lambda_i$ matrices of the $SU(3)$ symmetry
group. We show that the elements of the Jordan-Schwinger map are the
constants of motion of the quantum 3D isotropic harmonic oscillator.
Also, we show that the generalized Stokes operators together with
the $\lambda_i$ matrices may be used to expand the polarization
matrix. In section 3, by taking the expectation value of the
generalized quantum Stokes operators in a three-mode coherent state
of the electromagnetic field, we obtain the corresponding
generalized classical Stokes parameters. In section 4, by means of
the classical constants of motion of the 3D isotropic harmonic
oscillator we describe the geometrical properties of the
polarization ellipse. Finally, in section 5, we give some concluding
remarks.

\section{Jordan-Schwinger map and the harmonic oscillator constants of motion}

Usual classical Stokes parameters are defined as the expansion
coefficients of the polarization matrix \cite{MandelWolf,Simons} as
\begin{equation}
J_{2D}={1\over 2}\sum_{i=0}^{3}\sigma_i s_i.
\end{equation}
where $s_i$ are the four Stokes parameters, $\sigma_0={\bf
1}_{2\times 2}$ and $\sigma_i$, $i=1,2,3,$ are the Pauli matrices.
Since the $\sigma_i$ matrices are such that
$Tr(\sigma_i\sigma_j)=2\delta_{ij}$ and $Tr(\sigma_0\sigma_j)=0$,
then
\begin{equation}
Tr(J_{2D}\sigma_j)=s_j.
\end{equation}

\subsection{Usual quantum Stokes Operators}

The usual Stokes operators for a quantized plane electromagnetic
wave that propagates along the $z$ axis are defined as \cite{Jauch}
\begin{eqnarray}
S_0 = a^{\dag}\sigma_{0}a=a^\dag_1a_1+a^\dag_2a_2,\hspace{5ex}
S_1 = a^{\dag}\sigma_{1}a=a^\dag_1a_2+a^\dag_2a_1,\hspace{5ex}\nonumber \\
S_2 = a^{\dag}\sigma_{2}a=i(-a^\dag_1a_2+a^\dag_2a_1),\hspace{5ex}
S_3 = a^{\dag}\sigma_{3}a= a^\dag_1a_1-a^\dag_2a_2,\hspace{5ex}
\label{part}
\end{eqnarray}
where $a_j^\dag$ and $a_j$, $j=1,2,$ are the creation and
annihilation operators of the $j$-th  harmonic oscillator defined as
\begin{equation}
a_j^\dag={1\over \sqrt{2}}\left( x_j -i{p_j} \right),\hspace{3ex}
a_j={1\over \sqrt{2}}\left( x_j +i{p_j } \right),\label{mo1}
\end{equation}
with $[a_1,a_1^{\dag}]=[a_2,a_2^{\dag}]=1$ and
\begin{equation}
a^\dag=(a_1^\dag,a_2^\dag), \hspace{5ex} a=\pmatrix{a_{1}\cr
    a_{2}\cr}.
\end{equation}
We note that equations (\ref{part}) are a particular case of the
Jordan-Schwinger map with two kinematically independent bosons
\cite{Bieden}.

In the rest of the paper, the following observation is of
fundamental importance. The quantities (\ref{part}) are nothing more
than the constants of motion of the 2D isotropic harmonic
oscillator, with Hamiltonian $H_{2D}=a_1^\dag a_1+a_2^\dag a_2+1$.
In fact, we can show that
\begin{equation}
[S_i,H_{2D}]=0, \hspace{2ex} i=0,1,2,3.
\end{equation}
The commutation relations of the Stokes operators are immediately
obtained from the properties of the Jordan-Schwinger map
\cite{Bieden}. This leads us to the $SU(2)$ Lie algebra
\begin{equation}
\left[{S_\ell\over 2},{S_m\over 2}\right]=i\epsilon_{\ell
mn}{S_n\over 2},\hspace{3ex}  \ell,m,n=1,2,3,
\end{equation}
where $\epsilon_{\ell mn}$ is the well known totally antisymmetric
tensor.

We note that the angular momentum and the energy minus the zero
point energy of the 2D isotropic harmonic oscillator are equal to
\begin{equation}
L_z=S_2,\hspace{2ex} H_{2D}-1=S_0, \label{spin2D}
\end{equation}
respectively. According to Jauch and Rorhlich \cite{Jauch}, the spin
of the photon is given by $S_2$ and it is along the direction of
propagation. Therefore, the first equality in (\ref{spin2D}) means
that the angular momentum of the 2D isotropic harmonic oscillator is
equal to the spin operator of the photon.

Using equations (\ref{part}), we can write the polarization matrix
in terms of the constants of motion of the 2D isotropic harmonic
oscillator (usual quantum Stokes operators) as
\begin{equation}
J_{2D}={1\over 2 }\pmatrix{\langle S_0\rangle_\alpha+ \langle
S_3\rangle_\alpha&\langle S_1\rangle_\alpha +i\langle
S_2\rangle_\alpha\cr
    \langle S_1\rangle_\alpha-i\langle S_2\rangle_\alpha&
\langle S_0\rangle_\alpha-\langle S_3\rangle_\alpha\cr}
\end{equation}
where $\langle S_i\rangle_\alpha$ means the classical limit of the
Stokes operators, which will be found in section 3 by taking their
expectation values when the states of the electromagnetic field are
expressed as coherent or semiclassical states.

The physical and geometrical implications of the equality between
the Stokes operators and the constants of motion of the 2D isotropic
harmonic oscillator are extensively discussed in Ref. \cite{Mota}.

\subsection{Generalized Stokes Operators}

When the direction of arrival from the source is unknown \it a
priori\rm, we generalize the quantum Stokes operators as follows. By
using the Gell-Mann and Ne'eman $\lambda_i$ matrices of the $SU(3)$
symmetry group \cite{Gell-Mann} and the triplet of independent
harmonic oscillators (three independent bosons)
$a^\dag=(a_1^\dag,a_2^\dag,a^\dag_3$), we define the generalized
quantum Stokes operators as the Jordan-Schwinger map
$\Sigma_i=a^{\dag}\lambda_{i}a$, which explicitly are given by

\begin{eqnarray}
\Sigma_0 =a^\dag_1 a_1+a^\dag_2 a_2+a^\dag_3 a_3,\hspace{2ex}
\Sigma_1 = a^\dag_1a_2+a^\dag_2a_1,\hspace{2ex}
\Sigma_2 = i(-a^\dag_1a_2+a^\dag_2a_1),\nonumber \\
\Sigma_3 = a^\dag_1a_1-a^\dag_2a_2,\hspace{3ex} \Sigma_4 =
a^\dag_1a_3+a^\dag_3a_1,\hspace{4ex}
\Sigma_5 = i(a^\dag_3a_1-a^\dag_1a_3),\\
\Sigma_6 = a^\dag_2a_3+a^\dag_3a_2,\hspace{2ex} \Sigma_7 =
i(a^\dag_3a_2-a^\dag_2a_3),\hspace{2ex} \Sigma_8 = {1 \over
\sqrt{3}}(a^\dag_1a_1+a^\dag_2a_2-2a^\dag_3a_3),\nonumber \label{sg}
\end{eqnarray}
where we have used $\lambda_0={\bf 1}_{3\times 3}$.

From the commutation relations
$[a_1,a_1^{\dag}]=[a_2,a_2^{\dag}]=[a_3,a_3^{\dag}]=1$, we show that
the generalized quantum Stokes operators are the constants of motion
of the 3D isotropic harmonic oscillator with Hamiltonian
$H_{3D}=a_1^\dag a_1+a_2^\dag a_2+a_3^\dag a_3+{3\over 2}$, \it i.
e. \rm
\begin{equation}
[\Sigma_i,H_{3D}]=0,\hspace{2ex}i=0,...,8.
\end{equation}
Also, by the properties of the Jordan-Schwinger map \cite{Bieden},
we show that generalized quantum Stokes operators commutation rules
satisfy the $SU(3)$ Lie algebra
\begin{equation}
\left[{\Sigma_\ell\over 2},{\Sigma_m\over 2}\right]=i f_{\ell m
n}{\Sigma_n\over 2}, \hspace{2ex} \ell,m,n=1,...,8
\end{equation}
where the structure constants $f_{\ell m n}$ are totally
antisymmetric under exchange of any two indices and are given by
\begin{eqnarray}
f_{123}=1,\hspace{6ex} f_{147}= {1\over 2},\hspace{4ex}
f_{156}=-{1\over 2},\nonumber \\
f_{246}= {1\over 2},\hspace{6ex} f_{257}= {1\over 2},\hspace{6ex}
f_{345}={1\over 2},\\
f_{367}=-{1\over 2},\hspace{5ex} f_{458}={\sqrt{3}\over
2},\hspace{2ex} f_{678}={\sqrt{3}\over 2}.\nonumber
\end{eqnarray}

A careful analysis leads us to show that the angular momentum
operator $\hat {\bf L}=\hat {\bf r}\times\hat{\bf p}$ as well as the
energy operator of the 3D isotropic harmonic oscillator are
contained in the generalized quantum Stokes operators. Explicitly,
we can show that
\begin{equation}
L_1=\Sigma_7, \hspace{1ex} L_2=-\Sigma_5, \hspace{1ex}L_3=\Sigma_2,
\hspace{4ex} H_{3D}-{3\over 2}=\Sigma_0.
\end{equation}

Because of the first three equalities of this equation, the
generalization of what we have written after equation (\ref{spin2D})
means that the angular momentum of the 3D isotropic harmonic
oscillator essentially equal is to the spin operator of the photon.

We generalize the definition of the polarization matrix as follows
\begin{equation}
J_{3D}={1 \over 3}\lambda_0\langle\Sigma_0\rangle_\alpha +{1\over
2}\sum_{i=1}^{8}\lambda_i\langle\Sigma_i\rangle_\alpha,
\label{3dpola}
\end{equation}
again, $\langle \Sigma_i\rangle_\alpha$ is the classical limit of
the generalized quantum Stokes operators, which as it will be shown
in the next section, are the expectation values of the operators
$\Sigma_i$ in a coherent state of the electromagnetic field.

Since the $\lambda_i$ matrices are such that
$Tr(\lambda_i\lambda_j)=2\delta_{ij}$ and
$Tr(\lambda_0\lambda_i)=0$, $i,j=1,...,8$, then
\begin{equation}
Tr(J_{3D}\lambda_j)={1 \over 2}
Tr\left(\sum_{i=1}^8\lambda_i\lambda_j\langle\Sigma_i\rangle_\alpha\right)
=\langle\Sigma_j\rangle_\alpha. \label{traza3D}
\end{equation}
By using equations (10) and (15) the polarization matrix in terms of
the 3D isotropic harmonic oscillator constants of motion (or the
generalized quantum Stokes operators) takes the form
\begin{equation}
J_{3D}=\pmatrix{ {1\over 3}\langle\Sigma_0\rangle_\alpha+ {1\over
2}\langle\Sigma_3\rangle_\alpha+{1\over
2\sqrt{3}}\langle\Sigma_8\rangle_\alpha &{1\over
2}\langle\Sigma_1\rangle_\alpha-i{1\over
2}\langle\Sigma_2\rangle_\alpha &{1\over
2}\langle\Sigma_4\rangle_\alpha-i{1\over
2}\langle\Sigma_5\rangle_\alpha\cr {1\over
2}\langle\Sigma_1\rangle_\alpha+i{1\over
2}\langle\Sigma_2\rangle_\alpha &{1\over
3}\langle\Sigma_0\rangle_\alpha-{1\over
2}\langle\Sigma_3\rangle_\alpha +{1\over
2\sqrt{3}}\langle\Sigma_8\rangle_\alpha &{1\over
2}\langle\Sigma_6\rangle_\alpha-i{1\over
2}\langle\Sigma_7\rangle_\alpha\cr {1\over
2}\langle\Sigma_4\rangle_\alpha+i{1\over
2}\langle\Sigma_5\rangle_\alpha &{1\over
2}\langle\Sigma_6\rangle_\alpha+i{1\over
2}\langle\Sigma_7\rangle_\alpha &{1\over
3}\langle\Sigma_0\rangle_\alpha-{1\over
\sqrt{3}}\langle\Sigma_8\rangle_\alpha\cr}. \label{3dexplipola}
\end{equation}
We observe that the $\lambda_0$ coefficient in equation (15) is such
that equation (17) reduces to $J_{2D}$ when the propagation
direction of the plane electromagnetic wave is selected to be along
the $z$ axis. Also, we note that our definitions of $J_{3D}$ is such
that the trace of $J_{2D}$ and  $J_{3D}$ remains invariant.

It is important to note that the polarization matrix (\ref{3dpola})
formally can be defined for purely quantum states. This means that
it can be defined without taking the expectation values in a
semiclassical state of the electromagnetic field of the Stokes
operators $\lambda_i$. In this way, equation (\ref{traza3D}) becomes
to $Tr(J_{3D}\lambda_j)=\Sigma_j$. However, the implications of this
definition are out of the scope of this work.

\section{Generalized classical Stokes parameters}

We will obtain the classical limit for the generalized quantum
Stokes operators. To do this, we proceed as in Ref. \cite{Ruso} to
obtain the classical limit of the usual Stokes operators by taking
the expectation value of the operators (\ref{part}) in a two-mode
coherent state of the electromagnetic field. In our case, we compute
the mean value of the generalized quantum Stokes operators (10) in a
three-mode coherent state of the electromagnetic field
\begin{equation}
|\alpha_1,\alpha_2,\alpha_3\rangle=\sum_{n_1,n_2,n_3=0}^\infty{\alpha_1^{n_1}\alpha_2^{n_2}
\alpha_3^{n_3}\over \sqrt{n_1!n_2!n_3!}}|n_1,n_2,n_3\rangle.
\end{equation}
This leads us to the generalized classical Stokes parameters
\begin{eqnarray}
\langle\Sigma_0\rangle_\alpha =
|\alpha_{01}|^2+|\alpha_{02}|^2+|\alpha_{03}|^2,\hspace{4ex}
\langle\Sigma_1\rangle_\alpha = 2|\alpha_{01}||\alpha_{02}|\cos\Delta_{21},\nonumber\\
\langle\Sigma_2\rangle_\alpha =
2|\alpha_{01}||\alpha_{02}|\sin\Delta_{21},\hspace{13ex}
\langle\Sigma_3\rangle_\alpha = |\alpha_{01}|^2-|\alpha_{02}|^2,\nonumber\\
\langle\Sigma_4\rangle_\alpha =
2|\alpha_{01}||\alpha_{03}|\cos\Delta_{31},\hspace{8ex}
\langle\Sigma_5\rangle_\alpha = 2|\alpha_{01}||\alpha_{03}|\sin\Delta_{31},\\
\langle\Sigma_6\rangle_\alpha =
2|\alpha_{02}||\alpha_{03}|\cos\Delta_{32},\hspace{8ex}
\langle\Sigma_7\rangle_\alpha = 2|\alpha_{02}||\alpha_{03}|\sin\Delta_{32},\nonumber\\
\langle\Sigma_8\rangle_\alpha =
|\alpha_{01}|^2+|\alpha_{02}|^2-2|\alpha_{03}|^2,\hspace{20ex}
\nonumber \label{funda1}
\end{eqnarray}
where $\alpha_i=|\alpha_{0i}|\exp(i\phi_i)$ and $\Delta_{ij}\equiv
\phi_i-\phi_j$ is the classical phase difference.

As we will show in the next section, equation (19) represents the
Stokes parameters for three classical oscillations of amplitudes
$|\alpha_{0i}|$ and phases $\phi_i$. It is immediate to note that
these equalities reduce to the usual classical Stokes parameters as
a particular case when the amplitude and phase of the third
oscillation vanishes.

\section{Classical Stokes parameters, classical 3D isotropic harmonic oscillator constants of motion and the geometric properties of the polarization ellipse}

The classical three-dimensional isotropic harmonic oscillator is a
particle  that moves under the force
\begin{equation}
{\bf F} = -{\bf r}.
\end{equation}
By solving the Newton's second law and imposing the initial
conditions ${\bf r}_{t=0}=\bf{x}_o$ and ${\bf v}_{t=0}=\bf{v}_o$ we
obtain the solutions
\begin{equation}
x_i=a_i\cos{t}+b_i\sin{t},\hspace{2ex} i=1,2,3 \label{para}
\end{equation}
where $a_i=x_{oi}$ and $b_i=v_{oi}$. It is easy to see that these
solutions satisfy the ellipsoid equation
\begin{eqnarray}
&&x_1^2(a_2^2+b_2^2+a_3^2+b_3^2)+x_2^2(a_1^2+b_1^2+a_3^2+b_3^2)
+x_3^2(a_2^2+b_2^2+a_1^2+b_1^2)\nonumber \\
&&-2x_1x_2(a_1a_2+b_1b_2)-2x_2x_3(a_2a_3+b_2b_3)-2x_1x_3(a_1a_3+b_1b_3)\\
&&=(a_1b_2-a_2b_1)^2+(a_2b_3-a_3b_2)^2+(a_3b_1-a_1b_3)^2.\nonumber
\label{trayec}
\end{eqnarray}
This means that the orbit of the classical 3D isotropic harmonic
oscillator is contained in the ellipsoid. Moreover, since the
classical 3D isotropic harmonic oscillator potential has spherical
symmetry, its orbit is restricted to be in the perpendicular plane
to the classical angular momentum ${\bf L}_{cl}= \bf{r\rm \times \bf
p}$. Thus, the elliptic orbit of the classical 3D isotropic harmonic
oscillator is given by the intersection curve of the ellipsoid (22)
and the orthogonal plane to ${\bf L}_{cl}$, which contains the
origin of coordinates.

Equation (\ref{para}) can be written in oscillation form as
\begin{equation}
x_i=|\alpha_{0i}|\sin(\omega t +\phi_i),
\end{equation}
with
\begin{equation}
a_i=|\alpha_{0i}|\sin\phi_i,\hspace{5ex}b_i=|\alpha_{0i}|\cos\phi_i.
\end{equation}
These equalities imply that
\begin{equation}
\alpha_{0i}^2=a_i^2+b_i^2,\hspace{3ex}\sin\phi_i={a_i\over
\sqrt{a_i^2+b_i^2}}, \hspace{3ex}\cos\phi_i={b_i\over
\sqrt{a_i^2+b_i^2}}. \label{amplifase}
\end{equation}
The amplitudes and phases of the three classical oscillations of
equation (23) depend on the initial conditions $a_i$ and $b_i$
according to equations (25). Thus, if we substitute equation (25)
into equation (19), we incorporate the initial conditions in the
generalized classical Stokes parameters (constants of motion of the
classical 3D isotropic harmonic oscillator). In particular, at
$t=0$, the constant of motion of the angular momentum vector is
\begin{equation}
{\bf L}_{cl}=(\langle\Sigma_7\rangle_\alpha,
-\langle\Sigma_5\rangle_\alpha, \langle\Sigma_2\rangle_\alpha)=\bf
a\rm \times\bf b,
\end{equation}
where $\bf a\rm=(a_1,a_2,a_3)$, $\bf b\rm=(b_1,b_2,b_3)$, and the
ellipsoid equation (22) results to be
\begin{eqnarray}
&&x_1^2\left(
{4\langle\Sigma_0\rangle_\alpha-\langle\Sigma_8\rangle_\alpha\over
6}-{1\over 2}\langle\Sigma_3\rangle_\alpha\right)+
x_2^2\left( {4\langle\Sigma_0\rangle_\alpha-\langle\Sigma_8\rangle_\alpha\over 6}+{1\over 2}\langle\Sigma_3\rangle_\alpha\right)\nonumber\\
&&+x_3^2\left({2\langle\Sigma_0\rangle_\alpha+\langle\Sigma_8\rangle_\alpha\over
3}\right) -x_1x_2\langle\Sigma_1\rangle_\alpha-
x_2x_3\langle\Sigma_6\rangle_\alpha-
x_1x_3\langle\Sigma_4\rangle_\alpha\nonumber\\
&&={1\over 4}
\left(\langle\Sigma_7\rangle_\alpha^2+\langle\Sigma_5\rangle_\alpha^2
+\langle\Sigma_2\rangle_\alpha^2\right). \label{trayec2}
\end{eqnarray}
Following the definition of the Euler angles of Ref.
\cite{Goldstein}, we perform a rotation such that the direction of
the new $x_1$ axis coincide with that of the line of nodes, and the
direction of the new $x_3$ axis coincide with that of ${\bf
L}_{cl}$. The direction of the line of nodes (direction of the
intersection line between the orbit and the $x_1-x_2$ plane) is
found by a unitary vector in the $x_1-x_2$ plane, perpendicular to
${\bf
L}_{cl}=(\langle\Sigma_7\rangle_\alpha,-\langle\Sigma_5\rangle_\alpha,
\langle\Sigma_2\rangle_\alpha)$. This leads us to
\begin{eqnarray}
\sin\phi=n_x=\pm {\langle\Sigma_7\rangle_\alpha \over
\sqrt{\langle\Sigma_7\rangle_\alpha^2+\langle\Sigma_5\rangle_\alpha^2}},\nonumber\\
\cos\phi=n_y=\mp {\langle\Sigma_5\rangle_\alpha \over
\sqrt{\langle\Sigma_7\rangle_\alpha^2+\langle\Sigma_5\rangle_\alpha^2}}.
\end{eqnarray}
The orthogonality between ${\bf L}_{cl}$ \rm and the ellipse plane
leads to
\begin{equation}
\cos\theta={\langle\Sigma_2\rangle_\alpha\over
\sqrt{\left(\langle\Sigma_7\rangle_\alpha^2+\langle\Sigma_5\rangle_\alpha^2
+\langle\Sigma_2\rangle_\alpha^2\right)}}.
\end{equation}
On the other hand, it is well known that the constants of motion of
the classical 3D isotropic harmonic oscillator, in addition to the
energy and the angular momentum are given by the symmetric
Runge-type tensor \cite{Fradkin}
\begin{equation}
A_{ij}={1\over 2}(p_ip_j+\omega^2x_ix_j),\hspace{2ex}i,j=1,2,3
\end{equation}
It can be shown that the contraction of this equation with the
components of ${\bf L}_{cl}$ \rm yields to zero. This means that all
the geometric characteristics of the orbit must be determined by
$A_{ij}$. In fact, we can show that
\begin{eqnarray}
2A_{11}={2\langle\Sigma_0\rangle_\alpha+\langle\Sigma_8\rangle_\alpha
\over 6} +{\langle\Sigma_3\rangle_\alpha\over 2},\hspace{2ex}
2A_{22}={2\langle\Sigma_0\rangle_\alpha+\langle\Sigma_8\rangle_\alpha
\over 6}
-{\langle\Sigma_3\rangle_\alpha\over 2},\nonumber \\
2A_{33}={\langle\Sigma_0\rangle_\alpha
-\langle\Sigma_8\rangle_\alpha\over 3},\hspace{6ex} 2A_{12}={1\over
2}\langle\Sigma_1\rangle_\alpha,\hspace{6ex}
2A_{13}={1\over 2}\langle\Sigma_4\rangle_\alpha,\\
2A_{23}={1\over
2}\langle\Sigma_6\rangle_\alpha.\hspace{20ex}\nonumber
\end{eqnarray}
This shows that the geometric properties of the polarization ellipse
exclusively depend on the generalized classical Stokes parameters.
It can be shown that the eigenvalues of $A_{ij}$ depend only on the
energy and the magnitude of the angular momentum of the classical 3D
isotropic harmonic oscillator \cite{Fradkin}. $A_{ij}$ has an
eigenvector in the direction of the angular momentum, and the other
two eigenvectors are in the directions of the minor and major axis
of the elliptical orbit \cite{Fradkin}. Also, the eigenvectors of a
symmetric rank two tensor are determined by its eigenvalues as well
as its components \cite{Borisenko}. All this leads us to conclude
that the principal axis directions on the elliptical orbit are
completely determined by the generalized classical Stokes
parameters. Also, since ${\bf L}_{cl}$ is orthogonal to the
polarization ellipse, then ${\bf L}_{cl}$ points along the
propagation direction of the electromagnetic wave.

\section{Concluding Remarks}

This work links the quantum optics to classical optics by means of
quantum mechanics and it is a useful extension of the generalized
classical Stokes parameters into the quantum domain.

Although already there are treatments of the classical Stokes
parameters in the case of an \it a priori \rm unknown of the
electromagnetic propagation \cite{Roman, Suizos}, our treatment
results to be novel in the following aspects. We have introduced a
generalization of the quantum Stokes parameters of Jauch \it et. al.
\rm \cite{Jauch} using the Jordan-Schwinger map, three independent
bosons and the Gell-Mann and Ne'eman $SU(3)$ symmetry group
matrices. It was shown that the generalized quantum Stokes operators
result to be the expansion coefficients of the polarization matrix
in terms of the Gell-Mann and Ne'eman $SU(3)$ matrices. The
semiclassical limit of the generalized Stokes operators were
achieved by taking their expectation values in a three-mode coherent
state of the electromagnetic field. We have shown that the resulting
generalized classical Stokes parameters are consistent with the
generalized classical Stokes parameters which have recently been
published \cite{Suizos}. Thus, our treatment is more general than
those of references \cite{Roman, Suizos}, which are restricted to
the classical aspects of electromagnetic polarization.

We described by means of the classical 3D isotropic harmonic
oscillator constants of motion the geometrical properties of the
polarization ellipse. Particularly, we showed that the ellipsoid
coefficients and the symmetric Runge-type tensor of the classical 3D
isotropic harmonic oscillator are completely determined by the
generalized classical Stokes parameters. Also, we showed that the
first two Euler angles are intimately related to the components of
the orbital angular momentum of the classical 3D isotropic harmonic
oscillator.

Finally, we emphasize that the number of independent generalized
classical Stokes parameters are six. This is because in going from
(22) to (\ref{trayec2}), all of them were written in terms of the
six parameters, $a_i$ and $b_i$, $i=1,2,3$ which contain the initial
conditions of the classical 3D isotropic harmonic oscillator.

\section*{Acknowledgments}

R D Mota would like to thank the Departamento de Matem\'aticas del
Centro de Investigaci\'on y Estudios Avanzados del IPN where he was
a visitor during the preparation at this work.

The authors would like to thank to the referees for their comments
and suggestions that help us to give the final form to this work.

This work was partially supported by CONACyT Grant Number 37296-E,
SNI-M\'exico, COFAA-IPN, EDI-IPN, EDD-IPN and CGPI-IPN number
project 20000930.


\begin{thebibliography}{99}

\bibitem{Shurcliff1} Shurcliff  W A 1966 {\it Polarized Light: Production and Use}
(Harvard University Press,Cambridge, MA)

\bibitem{Shurcliff2} Ramachandran G N and Ramaseshan S 1961 {\it Encyclopedia of Physics} Ed. Fl\"ugge S
(Springer-Verlag, Germany)

\bibitem{Shurcliff3}Agarwal G S and  Chaturvedi S, 2003 {\it Journal of Modern Optics} \bf 50, \rm 711

\bibitem{Shurcliff4}Lehner J, Leonhardt U and Paul H 1996 {\it Phys. Rev. A} \bf 53, \rm 2727

\bibitem{Shurcliff5}Abouraddy A F, Sergienko A V, Saleh B E A and Teich M C,  2002 {\it Opt. Commun.} {\bf 201} 93

\bibitem{Shurcliff6} Jaeger G, Teodorescu-Frumosu M, Sergienko A V, Saleh B E A and Teich M C  2003 {\it Phys. Rev. A} \bf 67, \rm 032307

\bibitem{Stokes} Stokes G G 1852 \it{Trans Cambridge Philos. Soc.} \rm\bf{9} \rm 399
(Reprinted in Stokes G G  1966 \it{Mathematical and Physical Papers}
\rm(Johnson Reprint Corporation, N Y and London))

\bibitem{Wiener} Wiener N 1930 {\it Acta Math.} {\bf 55}, 117
(Reprinted in Wiener N 1964 {\it Generalized Harmonic Analysis and
Tauberian Theorems} (The M. I. T. Press, USA))

\bibitem{Fano}Fano U 1954 {\it Phys. Rev.}\bf 93,\rm 121

\bibitem{Jackson}Jackson J D 1975 {\it Classical Electrodynamics} (John Wiley \& Sons Inc., USA) p 273

\bibitem{Jauch} Jauch J M and Rohrlich F 1976 {\it The Theory of Photons and Electrons} (Springer-Verlag, Berlin) p 41

\bibitem{Roman}Roman P 1959 {\it Nuovo Cimento} {\bf 13}, 974

\bibitem{Suizos} Carozzi T, Karlsson R and Bergman J 2000 \it{Phys. Rev. E} \rm\textbf{65} 2024.

\bibitem{MandelWolf} Mandel L and Wolf E 1995 {\it Optical Coherence and Quantum Optics}
(Cambridge University Press, USA) p 349

\bibitem{Simons} Simmons J W and Guttmann M J 1970 {\it STATES, WAVES AND PHOTONS: A Modern
Introduction to Light} (Addison-Wiley Publishing Company, USA) p 74

\bibitem{Bieden} Biedenharn L C and Louck J D 1981 {\it Angular Momentum in Quantum Physics} (Addison-Wesley Publishing Company, USA) p 213

\bibitem{Mota}Mota R D, Xicot\'encatl M A and Granados V D, 2003 To be published

\bibitem{Gell-Mann} Gell-Mann M and  Ne'eman Y 1964 {\it The Eight-fold Way} (Benjamin, N Y, USA)

\bibitem{Wybourne} Wybourne B G 1974 {\it Classical Groups For Physicists} (John Wiley \& Sons, USA) p 268

\bibitem{Ruso} Tanas R and Gantsog T S 1992 {\it Opt. Commun.} {\bf 87} 369.

\bibitem{Goldstein} Goldstein H 1980 {\it Classical Mechanics} (Addison-Wesley Publishing Company, USA) p 143

\bibitem{Fradkin} Fradkin D M  1967 {\it Prog. Theor. Phys.} \rm \textbf{37} 798\\
 Fradkin D M  1965 \it Am. J. Phys. \rm \textbf{33} 207

\bibitem{Borisenko} Birisenko A I and Tarapov I E 1979 {\it Vector and Tensor Analysis}
(Dover Publications, Inc., USA) p 115

\end{thebibliography}
\end{document}